# Will NIF Work?

W. J. Nellis

Department of Physics

Harvard University

August 28, 2009

It is vital that new clean and abundant sources of energy be developed for the sustainability of modern society. Nuclear fusion of the hydrogen isotopes deuterium and tritium, if successful, might make a major contribution toward satisfying this need. The U.S. has an important effort aimed at achieving practical inertial confinement fusion, ICF, which has been under development for decades at the Lawrence Livermore National Laboratory. The National Ignition Facility (NIF) is a giant laser to multiply-shock and thus quasi-isentropically compress a capsule of deuterium-tritium (DT) to high density and temperature, where the fusion rate is proportional to density squared times temperature to the fourth power. The principal problem that must be solved for NIF to work successfully is elimination of the Rayleigh-Tailor (R-T) instability that originates from the interface between the solid shell and the DT fuel within it. The R-T instability poisons the fusion reaction by reducing the temperature of the DT achieved by dynamic compression driven by the NIF laser.

The primary technological problem today is one of Condensed Matter and Materials Physics (CMMP), rather than laser technology and plasma physics.

The CMMP of the fuel capsule that must be done to minimize growth of the R-T instability is yet to be done. <u>Based on what is known today, it is unlikely that NIF will produce practical amounts of fusion energy</u>.

**I. Introduction**

Scientists have realized for decades that clean non-carbonaceous sources of energy are required to meet the growing energy demand worldwide. In the 1990s the President's Committee of Advisors on Science and Technology (PCAST) concluded that "world electricity use in particular is likely to triple by the year 2050 [1]." The threat posed to world health, stability, and to its Economy is sufficiently great that several new energy sources are probably required to meet this challenge.

Because nuclear-fission reactors have been used successfully as a source of commercial power, it is virtually certain that an increasing number of nuclear-fission reactors will be put into operation in the coming years. Use of nuclear-fission power means that safe and efficient ways of dealing with nuclear waste must be developed.

In the longer term nuclear fusion might be able to help satisfy the growing energy demand, as well. Feynman [2] pointed out that "the energy that can be obtained from 10 quarts of water per second is equal to all the electrical power generated in the United States" and that it is "up to the physicists to figure out how." The two main attractions of fusion are the 14 MeV of kinetic energy that is released in each fusion of a deuteron and a triton and the virtually unlimited amount of deuterium that exists in the oceans. Tritium is made from deuterium by neutron irradiation. Fusion fuel is available in virtually unlimited quantities. Despite its great promise fusion is yet to be realized as a commercial source of energy. *The purpose of this white paper is to use the published literature to review the physics of Inertial Confinement Fusion, including its prime impediment (which was published in 1972), present status, and likelihood of success. Suggestions will be made for Condensed Matter and Materials Physics (CMMP) [3] research that must be done to increase the likelihood of the success of NIF.*

The plasma physics of laser-driven ICF has been described by Lindl [4]. The fuel capsule is a spherical solid shell that contains a sphere of condensed DT. Pulsed laser beams from NIF (pulse widths of tens of ns) rapidly generate a plasma on the outer surface of the fuel capsule, which will ablate that surface and generate a converging strong shock in the solid shell containing the DT fuel. DT is used because cross sections for fusion of deuterons and tritons (DT) are 3 orders of magnitude larger than for any other fusion process. This shock in the shell eventually breaks out of the solid shell and enters the DT. Interfacial instabilities caused by shock break out are called Richtmyer-Meshkov (R-M) instabilities [5,6]. In ICF R-M instabilities nucleate Rayleigh-Tailor (R-T) instabilities [7-8], which then grow under high accelerations of the shell during spherical convergence. R-T instabilities poison fusion.

High temperatures T and densities $\rho$ of DT are necessary for (i) positive ions to get over the potential barrier due to electrons between nuclei and (ii) to have a sufficiently high density of DT to produce substantial fusion reactions. Fusion cross sections thermally averaged over a Maxwell-Boltzmann velocity spectrum ($\sigma_{th}v$) vary as $T^4$. Since the reaction rate R is largest for an equimolar mixture of deuterium and tritium, the fusion rate $R \sim \rho^2 T^4$.

To maximize compression by keeping the fuel cool hydrodynamically, several driving pulses are generated with a laser pulse stepped in time. The effect is to compress the DT quasi-isentropically, a process with no fundamental limit on $\rho$ and T. Ideally, on convergence to the center, $\rho$s and Ts are maximum and DT fuses, which results in 14-Mev neutrons and 3.5-Mev alpha particles. A 3.5-Mev alpha particle has a short mean-free path, which produces more local heating facilitating more fusion. In a fusion reactor kinetic energy of the neutron would be converted into heat and probably used to fission $U^{235}$, which would produce more heat. Heat would be converted into electrical power.

Idealized calculations at Lawrence Livermore National Laboratory (LLNL) in 1972 showed central conditions in DT in the fuel capsule were expected to reach $\rho$s as high at ~1,000 gcm$^{-3}$ at Ts of ~10 keV [9]. Unfortunately, central temperatures are substantially less than ideal because of interfacial Rayleigh-Tailor (R-T) instabilities between solid shell and DT fuel and the non-ideality of spherical convergence.

**II. New Facility for ICF (~$5 B) is Virtually Complete: CMMP Research Needed**

For the past 13 years the National Ignition Facility (NIF) has been under construction at LLNL for large-scale ICF. Over the past 40 years LLNL has spent an estimated $10 B (in today's $) on ICF, well over half of which has been spent on NIF construction costs. The size of NIF is roughly a cubic football field; the size of the fuel capsule, whose performance determines NIF's fusion production, is of order a cm. Despite the financial and human resources and time spent on NIF, the key condensed matter and materials physics issues of the fuel capsule remain unsolved and the R-T instability continues to be the limiting feature of NIF performance.

### III. Crucial CMMP Issues for ICF

While both the physics of R-T growth and the equations of state of DT (hydrogen) and shell material must be known, the R-T instability is by far the major issue and is the one focused on in this discussion. The challenge comes from the fact that, as in ICF, R-T experiments on solids need to be done under dynamic (~10 ns) compression at shock breakout pressures in the shell of several 100 GPa (several Mbar). The short time scales and high loading rates on potential solid capsule materials are such that the nature of their mechanical response must be determined experimentally. In this regime the dynamics of compression, strength of materials, phase transitions, and microstructural heterogeneous deformation have essentially never been measured experimentally and must be taken into account correctly to design fuel capsules. These responses cannot be calculated, as has been attempted for decades. We now illustrate the basic phenomenon.

Figure 1 is a schematic of a simple representative ICF target [4]. A laser pulse is incident simultaneously over the outer spherical plastic (CH) surface. A converging shock wave is generated by the pulse and the glass pusher is subsequently accelerated inward by the converging shock wave. In an actual ICF target, DT would fill the inner sphere.

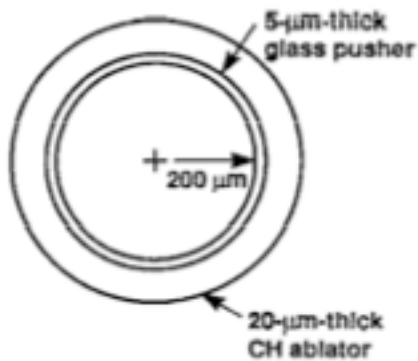

Fig. 1 Static representation of ICF target.

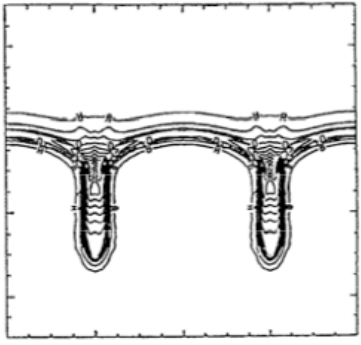

Fig. 2 Late-time experimental photograph of an initially planar Al foil (top) accelerated into vacuum (bottom) by fast laser pulse incident on top Al surface (not shown).

Figure 2 is a fast photograph of an initially planar Al foil that has been accelerated downward into vacuum by a shock wave that was generated by a fast laser pulse incident on the upper surface of the Al foil [4]. The Al spikes that have formed pointing into vacuum are a result of the R-M instability that grew when the shock in Al broke out into the vacuum. It is R-T spikes that grow from such R-M instabilities under high accelerations at later times that are the show-stopper of ICF.

Ideally both the inner surface of the fuel capsule and the outer surface of the DT fuel ball would remain perfectly spherical during the fast ICF implosion. In this case the temperatures and densities calculated by a simple 1-D computer code are roughly accurate to the extent that the equation-of-state (EOS) of the capsule and fuel are known accurately. These temperatures and densities could be used to calculate fusion burn rate and fusion energy released.

Unfortunately, the R-M instability and the R-T instability, which occurs at late times, destroy the integrity of the interface and mix the deuterium-tritium fuel with "high-Z" (C, Si, O, …) material from the capsule. When the high-Z atoms mix with the fuel heated by compression, thermal energy of the fuel ionizes atoms from the shell, which in turn lowers the temperature of the fuel, which lowers the fusion reaction rate ($R \sim T^4$).

The simplest mathematical description of these instabilities is

$$\eta(\tau)/\eta(0) = 1 + \gamma\tau, \qquad (1)$$

$$\gamma \sim [(\rho_2 - \rho_1)/(\rho_2 + \rho_1)]/\lambda, \qquad (2)$$

where $\eta(\tau)$ is the amplitude of a sinusoidal perturbation at the interface at time $\tau$, $\eta(0)$ is the amplitude at $\tau=0$, $\gamma$ is the lowest-order growth rate, $\rho_2$ and $\rho_1$ are the mass densities of shell and DT, and $\lambda$ is the wave length, or size, of the perturbation. The shortest wavelengths have the largest growth rates and must be avoided.

The concept of ICF was first proposed by Nuckolls et al [9]. They said, "The implosion of the (ICF) pellet by diffusion driven ablation-generated pressure is hydrodynamically stable, except for relatively long wavelength surface perturbations, which grow too slowly to be damaging." The "long wavelength surface perturbations" do grow very slowly under R-M and R-T. Unfortunately, short wavelength surface perturbations also exist in real systems and these grow sufficiently fast and large to substantially reduce fusion reactions. Quantitative understanding of R-M and R-T instability growth is needed for target design and is yet to be obtained.

Recently a paper entitled "A comparative study of the turbulent Rayleigh–Taylor instability using high-resolution three-dimensional numerical simulations: The Alpha-Group collaboration," was published by Los Alamos, Lawrence Livermore, Sandia National Laboratories, the University of Chicago, and other universities and national laboratories [10]. This article says that not even the lowest-amplitude initial growth rate of the R-T instability is understood quantitatively. The same must also be true for growth in later stages. Computational simulations for more than 35 years have provided no insight into eliminating the R-T instability [11].

Based on Eqs. (1) and (2) it has been assumed that the R-M and R-T instabilities could be eliminated simply by fabricating the inner spherical surface of the capsule to be spherically smooth to a few nm. Since the inception of the ICF program in the early 1970s the fuel capsule was made from compressible plastics. Such plastics have no strength needed to impede instability growth. About the time it was shown that the R-T computational program had not been useful for this problem [10], the ICF Program began considering the strongest material known, diamond. Unfortunately for NIF, it appears that it does not matter how smooth capsule surfaces are fabricated because when the capsule is shocked by the laser pulse, rapid deformation induces defects in the shell, which show up as small interfacial imperfections on the interface when the shock breaks out of the shell. These shock-induced imperfections nucleate sites for growth of R-T instabilities under subsequent fast spherical convergence.

Hare et al at LLNL [12] took fast photographs as shock waves traveled through sapphire crystals in various directions in the sapphire lattice at several shock stresses. Some of their photos (negatives) are shown in Fig. 3. Light is emitted from the darker regions. These pictures show that failure in sapphire under shock compression depends on crystallographic orientation and shock stress. The heterogeneous hot spots in Fig. 3 probably originate from locations of brittle fracture and plastic slip. Such heterogeneous defects will probably act as nucleation sites of R-M instabilities because small shock-induced perturbations generated in the bulk will be imprinted on the interface at shock breakout. Subsequent acceleration will probably cause their R-T growth [13].

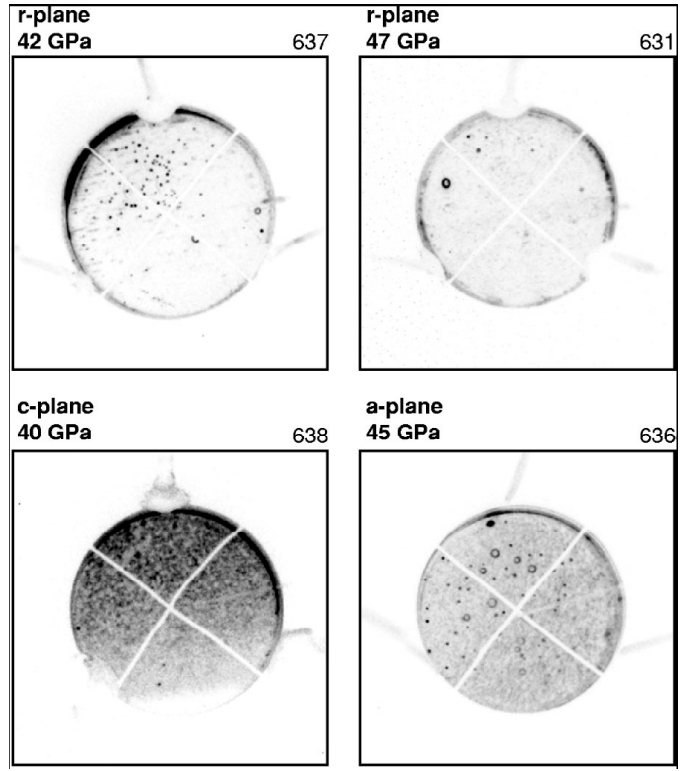

Fig. 3  Fast photographs (negatives) taken as shock wave of ~43 GPa traversed sapphire crystals of various orientations. Dark spots are sources of emitted light (After Hare et al 2002).

Those shock compression experiments on sapphire are expected to be representative of most if not all strong materials. One such material is diamond, which is even stronger than sapphire and expected to show similar phenomena. Recently, Celliers et al [14] from LLNL reported very nice measurements taken with a two-dimensional imaging VISAR (Velocity Interferometer System for Any Reflector) over an 800 micron field of view, which was sensitive to mode wavelengths on shock breakout ranging from 2.5 to 100 micron. At breakout of a multi-Mbar shock from polycrystalline diamond, the method showed "a high degree of non-uniformity on spatial scales of a few microns or less." Celliers et al have demonstrated that R-T growth rates can be large no matter how smooth the capsule is fabricated. They have also demonstrated that diamond is probably not a material that should be used in a fuel capsule. This means that the optimal material must be found from which to fabricate the shell. Something that is yet to be done.

Similar experiments showed that surface roughness could be reduced by going to shock stresses near melting or in the melt region of diamond. Those results indicate that reducing strength produces smoother surfaces. However, it is not surface roughness *per se* that poisons fusion. Rather, it is interfacial roughness and mixing that occur after subsequent R-T growth under high interfacial accelerations. Also, at sufficiently high shock pressures and weak shell material, the concept of R-T amplitude growth becomes meaningless. For a sufficiently strong shock on breakout, the impulse will simply spray bits of the shell into the DT, which will also poison fusion for the same reason the R-T instability does. Thermal energy will be absorbed by ionizing shell bits. A multi-Mbar shock is likely to do this. This might be what happened in the polycrystalline diamond sample. In a polycrystalline sample, it is probably the strength of grain boundaries that is the strength that maintains shell integrity on shock breakout and not the intrinsic strength of the grains themselves. The same consideration holds for a shell made of nanocrystalline Be, for example. It is unlikely that the integrity of a smooth surface of such a shell can be maintained on breakout of a multi-Mbar shock wave. It is very important to determine experimentally whether or not shell-surface integrity can be maintained on breakout of a multi-Mbar shock. Computations cannot determine this.

Suggestions: i) Implosion of a solid ICF shell requires a material that is sufficiently strong to resist R-M and R-T growth and sufficiently weak not to generate too many nucleation sites for R-M and R-T growth. Such a material must be found from which to fabricate a smooth-surfaced shell in the fuel capsule. A large number of potential materials with varying strengths, single crystals and polycrystalline, should be investigated to identify optimal ones. ii) For potential shell materials, single crystals and polycrystalline, find the maximum stress that the material can withstand without "too much surface ejecta," which would poison fusion. This pressure is a design limit on the first shock in the shell.